\documentclass[letter]{aa} 
\usepackage{graphicx}
\usepackage{epsfig}
\usepackage{psfig}
\usepackage{txfonts}
\usepackage{natbib}
\bibpunct{(}{)}{;}{a}{}{,} 

\def\ccsnrate{\mbox{\,$\nu_{\rm CCSN}$}}

\def\asec{\ifmmode ^{\prime\prime}\else$^{\prime\prime}$\fi}

\def\degs{\ifmmode ^{\circ}\else$^{\circ}$\fi}
\def\amin{\ifmmode ^{\prime}\else$^{\prime}$\fi}
\def\asec{\ifmmode ^{\prime\prime}\else$^{\prime\prime}$\fi}

\def\degs{\ifmmode ^{\circ}\else$^{\circ}$\fi}
\def\amin{\ifmmode ^{\prime}\else$^{\prime}$\fi}
\def\EE#1{\times 10^{#1}}

\def\cm{\mbox{\,cm}}

\def\cm3{\mbox{\,cm$^{-3}$}}

\def\ergs{\mbox{\,erg~s$^{-1}$}}
\def\ergshz{\mbox{~erg~s$^{-1}$~Hz$^{-1}$}}

\def\lsim{\!\!\!\phantom{\le}\smash{\buildrel{}\over
 {\lower2.5dd\hbox{$\buildrel{\lower2dd\hbox{$\displaystyle<$}}\over
                                 \sim$}}}\,\,}
\def\gsim{\!\!\!\phantom{\ge}\smash{\buildrel{}\over
{\lower2.5dd\hbox{$\buildrel{\lower2dd\hbox{$\displaystyle>$}}\over
                               \sim$}}}\,\,}

\begin{document}
 \title{An extremely prolific supernova factory in the buried nucleus of the starburst galaxy IC 694}

   \author{M.A. P\'erez-Torres \inst{1}
      \and 
      C. Romero-Ca\~nizales \inst{1}
      \and
      A. Alberdi \inst{1}
      \and
      A. Polatidis \inst{2,3}
                   }
          
   \institute{Instituto de Astrof\'{\i}sica de Andaluc\'{\i}a - CSIC, PO Box 3004, 18008 Granada,  Spain \\
             \email{torres@iaa.es}
         \and
             Joint Institute for VLBI in Europe (JIVE), Dwingeloo, The Netherlands
        \and
             ASTRON, Dwingeloo, The Netherlands
            }

\date{Received 23 July 2009; Accepted 17 October 2009}

\abstract
{The central kiloparsec of many local Luminous Infra-red Galaxies
are known to host intense bursts of massive star formation,
leading to numerous explosions of core-collapse supernovae (CCSNe).
However, the dust-enshrouded regions where those supernovae explode hamper their
detection at optical and near-infrared wavelengths.}
{We investigate the nuclear region of the starbust galaxy IC 694 (=Arp 299-A)
at radio wavelengths, aimed at discovering recently exploded CCSNe, 
as well as to determine their rate of explosion,  which carries crucial 
information on star formation rates, the initial mass function and starburst scenarios at work.}
{We use  the electronic European VLBI Network to image with milliarcsecond resolution 
the 5.0 GHz compact radio emission of the innermost nuclear region of IC 694.}
{Our observations reveal the presence of a rich cluster of 26 compact radio emitting sources in the central 150 pc of  the nuclear starburst in IC 694. The large brightness temperatures observed for the compact sources indicate a non-thermal origin for the observed radio emission, implying that most, if not all, of those sources are young radio supernovae (RSNe) and supernova remnants (SNRs). 
We find evidence for at least three relatively young, slowly-evolving,
long-lasting RSNe (A0, A12, and A15) which appear to display unusual
properties; suggesting that the conditions in the local circumstellar
medium (CSM) play a significant role in determining the radio
behaviour of expanding SNe.
Their radio luminosities are typical of normal RSNe, which result from the explosion of Type IIP/b and Type IIL SNe.  All these results yield support for a recent (less than 10$-$15 Myr) instantaneous starburst in the innermost regions of IC 694.  }
{}


   \maketitle
%

\section{Introduction}

The observed rate at which massive stars (M $\gsim$8 M$_{\odot}$) die as CCSNe, \ccsnrate, can be used as a direct measurement of the current star formation rate (SFR) in galaxies, and provides unique  information on the initial mass function (IMF) of massive stars.  While the rate at which stars die in normal galaxies is rather low (e.g., one SN is expected to explode in the Milky Way every $\sim$50 yr), the CCSN rate in Luminous and Ultraluminous Infra-red Galaxies (LIRGs, L$_{\rm IR} \equiv$ L$[8-1000\, \mu\,{\rm m}]\geq 10^{11}$ L$_{\odot}$; and ULIRGs, L$_{\rm IR} \geq 10^{12}$ L$_{\odot}$; \citet{sanders96}) is expected to be at least one or two orders of magnitude larger than in normal galaxies \citep{condon92}, and hence detections of SNe in (U)LIRGs offer a promising way of determining the current star formation rate in nearby galaxies.  

However, the direct detection of CCSNe in the extreme densities of the central few hundred pc of (U)LIRGs is extremely difficult, 
since emission in the
visual band suffers very significant extinction in those regions which
contain large amounts of dust, and can at best yield only an upper
limit to the true value of \ccsnrate.  Fortunately, it is possible to directly probe the star forming activity in the innermost regions of (U)LIRGs by means of high angular resolution ($\leq 0.05$ arcsec), high-sensitivity ($\leq$0.05 mJy) radio searches of CCSNe, as radio does not suffer from dust extinction, and the angular resolution yielded by current Very Long Baseline Interferometry (VLBI) arrays, of the order of a few milliarcsec at cm-wavelengths,  is able to detect individual radio supernovae at large distances in the local Universe.  

Starburst activity in the circumnuclear regions of (U)LIRGs implies both the presence of a high number of massive stars and a dense surrounding medium, so bright radio SNe are expected to occur \citep{chevalier82, chugai97}, and high-resolution radio observations have shown that highly-extinguished CCSNe do exist in the circumnuclear ($r\lsim$1~kpc) region of local (U)LIRGs \citep{smith98,lonsdale06,colina01,neff04,perez-torres07,kankare08}.  Therefore, VLBI observations can set strong constraints on the properties of star formation in the dust-enshrouded environments encountered in (U)LIRGs.

Arp~299 consists of two interacting galaxies (IC 694 and NGC 3690), which are in an early merger stage \citep{keel95}. At a luminosity distance of 44.8 Mpc \citep{fixsen96} for $H_0 = 73$~km~s$^{-1}$~Mpc$^{-1}$, Arp~299 has an infrared luminosity $L_{\rm IR} \approx 6.7\EE{11} L_{\odot}$ \citep{sanders03}, which is approaching the ULIRG category.  The innermost $\sim$150 pc nuclear region of Arp 299-A (see Figure \ref{fig,arp299}) is heavily dust-enshrouded,  thus making the detections of SNe very challenging even at near-infrared wavelengths. Yet, Arp 299 hosts recent and intense star forming activity, as indicated by the relatively high frequency of {supernovae discovered at optical and near-infrared wavelengths} in their outer, much less extinguished regions \citep{forti93,vanburen94,li98,yamaoka98,qiu99,mattila05}.  

The brightest component at infrared and radio wavelengths is IC 694 (A in the top panel of Figure \ref{fig,arp299}; hereafter Arp 299-A), accounting for $\sim$50\% of the total infrared luminosity of the system \citep{almudena00,charmandaris02}, and for $\sim$70\%\/ of its 5 GHz radio emission \citep{neff04}. Numerous H~II regions populate the system near star-forming regions, suggesting that star formation has been occurring at a high rate for the last $\sim$10 Myr \citep{almudena00}.  Given that IC~694 accounts for most of the infrared emission in Arp 299, it is the region that is most likely to reveal new SNe \citep{condon92}. Since optical and near-infrared observations are likely to miss a significant fraction of CCSNe in the innermost regions of Arp 299-A due to  large values of extinction [$A_V \sim 34-40$ \citep{gallais04,almudena09}] and the lack of the necessary angular resolution, radio observations of Arp 299-A at high angular resolution, high sensitivity are the only way of detecting new CCSNe and measuring directly and independently of models its CCSN and star formation rates.  In fact, Very Long Baseline Array (VLBA) observations carried out during 2002 and 2003 resulted in the detection of five compact sources \citep{neff04}, one of which (A0) was identified as a young SN.

\begin{figure}
\centering  
\includegraphics[width=90mm,angle=0]{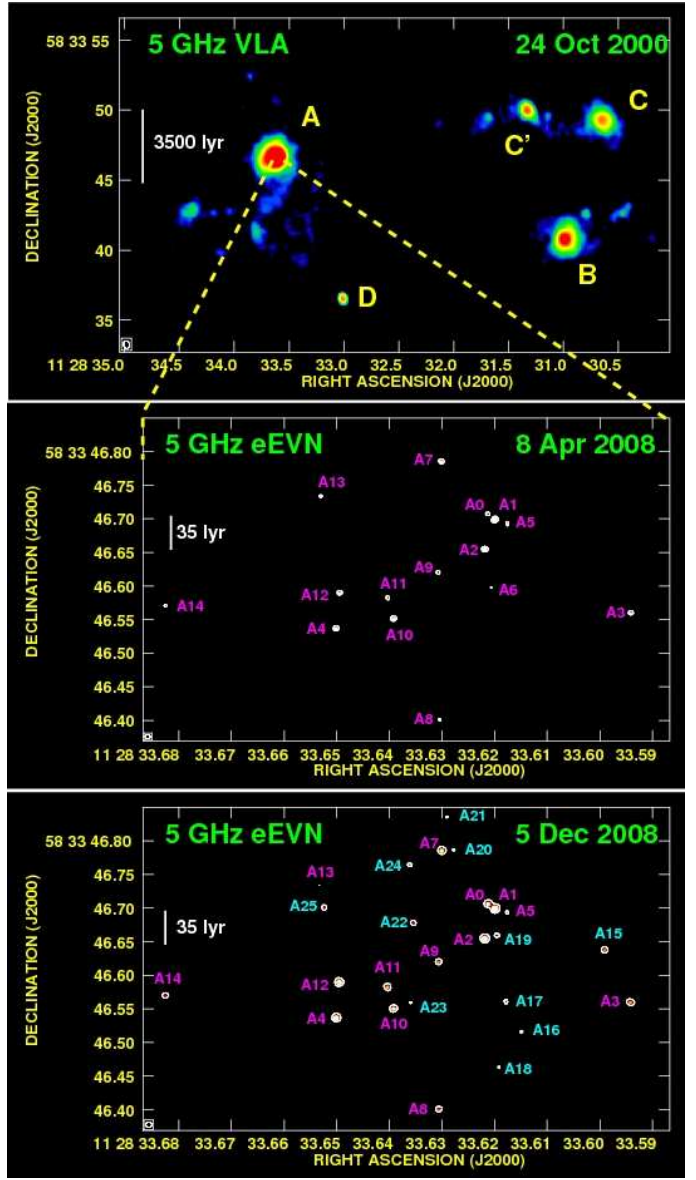} \\
\caption{ {\it
  Top:} 5 GHz VLA archival observations of Arp 299 on 24 October 2000,
displaying the five brightest knots of radio emission in this merging
galaxy. 
{\it  Middle and bottom:} Contour maps drawn at five times the r.m.s. of our 5 GHz eEVN observations of the central 500 light
years of the Luminous Infrared galaxy Arp 299-A on 8 April 2008 and 5
 December 2008, revealing a large population of relatively bright, compact, non-thermal emitting sources. The size of
 the FWHM synthesized interferometric beam was of (0.6 arcsec $\times$
 0.4 arcsec) for the VLA observations, and of (7.3 milliarcsec $\times$
 6.3 milliarcsec) and (8.6 milliarcsec $\times$ 8.4 milliarcsec) for
 the EVN observations on 8 April 2008 and 5 December 2008,
 respectively.  To guide the reader's eye, we have shown in cyan the components
 detected only on the 5 December 2008 epoch.}
 \label{fig,arp299}
\end{figure}

\begin{figure}[h!]
\includegraphics[width=95mm,keepaspectratio=true,angle=0]{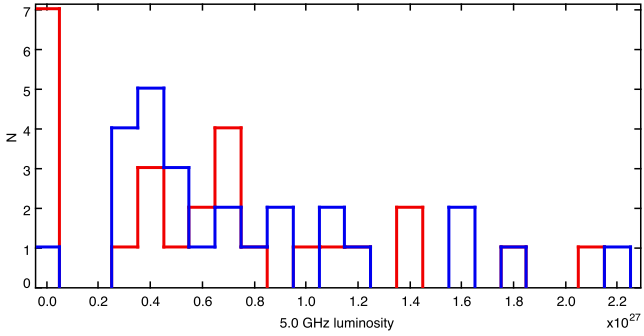}\\
\caption{5.0 GHz luminosity histogram for the VLBI components detected
on 8 April 2008 (red) and 5 December 2008 (blue). The values centered at 0.0
correspond to the upper limits quoted in Table~ \ref{tab,evn}. See main text for details.}
\label{f-lumin}
\end{figure}

\section{eEVN observations and results}

We used the electronic European VLBI Network (e-EVN) \citep{szomoru06} to image Arp~299-A at a frequency of 5~GHz over 2 epochs, aimed at directly detecting recently exploded core-collapse supernovae through the variability of their compact radio emission (see appendix \ref{app,evn} for a detailed explanation of our observing strategy, calibration and imaging procedures, and source detection and techniques for flux density extraction).  The attained off-source root-mean-square (r.m.s.) noise level was of 39 $\mu$Jy/beam and 24 $\mu$Jy/beam for the 8 April 2008 and 5 December 2008 observations, respectively, and reveal the existence of 26 compact components above 5 r.m.s. (see Figure \ref{fig,arp299}).  Since the EVN radio image on 5 December 2008 is much deeper than the one obtained on 8 April 2008, it is not surprising that we detected a larger number of VLBI sources in our second epoch (25) than in our first one (15). This allowed us to go back to our first-epoch image and extract the flux density for the new components (A15 through A25 in Figure \ref{fig,arp299}), and which show $\geq 5$r.m.s. detections only in the December 2008 image). This procedure allowed us to recover four components above 3$\sigma$ (A15, A18, A22, and A25), based on a positional coincidence with the peak of brightness of our second epoch of better than $\sim 0.5$ milliarcsec, i.e.,  much smaller than the synthesized interferometric beam).

Our results demonstrate the existence of a very compact rich nuclear starburst in Arp 299-A and, in general, are in excellent agreement with independent results reported by \citet{ulvestad09}. The angular size encompassed by the radio emitting sources in Arp 299-A is less than 0.7''$\times$0.4'', corresponding to a projected linear size of (150$\times$85)~pc.  To facilitate comparisons, we define here a fiducial supernova radio luminosity equal to three times the image r.m.s.  in the 8 April 2008 epoch, which corresponds to 2.9$\EE{26}$\ergshz. In this way, the radio luminosities for the VLBI components range between 1.1 (A25) and 7.3 (A1) and between 1.0 (A13) and 7.7 (A1) times the fiducial value, for the VLBI observations on 8 April 2008 and 5 December 2008, respectively (see Table~\ref{tab,evn} for details).

\section{Discussion}
The radio emission from the compact sources detected from our VLBI observations can be explained in principle within two different physical scenarios: (i) thermal radio emission from super star clusters (SSCs) hosting large numbers of young, massive stars that ionize surrounding H~II regions; (ii) non-thermal radio emission from supernova remnants (SNRs) and/or young radio supernovae (RSNe), i.e., recently exploded core-collapse supernovae where the interaction of their ejecta with their surrounding circumstellar or interstellar medium (CSM or ISM, respectively) would give rise to significant amounts of synchrotron radio emission.

The existence of SSCs in Arp~299-A has been demonstrated by their apparent detection using 2.2-$\mu$m adaptive optics imaging \citep{lai99}. Further evidence in this direction comes from {\it
  Hubble Space Telescope\/} ({\it HST}\/) FOC and NICMOS images, which reveal a population of young stellar clusters in the central regions of Arp 299 \citep{almudena00}.  The total 5.0 GHz radio luminosity in compact sources is 1.7$\EE{28}$\ergshz\/ and 2.0$\EE{28}$\ergshz\/ on 8 April 2008 and 5 December 2008, respectively.  
However, the high flux densities in Tablefor most of the compact sources in Arp 299-A ( $\leq$9 milliarcsec; see Table~\ref{tab,tb}), indicate brightness temperatures, $T_{\rm B}$, which largely exceed the thermal temperatures expected from SSCs ($\lsim 2\EE{4}$K), thus ruling out a thermal origin for the compact radio emission traced by our eEVN observations.

Therefore, the observed radio emission must be due to young radio supernovae, SNRs, or both.  We show in Table \ref{tab,evn} the flux densities and luminosities for all the components identified in our observations, and classify the objects according to their variability.  The majority of sources do not show any evidence for significant variability which is consistent with their identification as SNR.  Only three sources (A6, A12, and A15) show appreciable flux density variations between our two consecutive VLBI observations (see Table~ \ref{tab,evn} for details), which is very difficult to reconcile with their radio emission being due to SNRs.  A6, A12, and A15 are also detected in previous VLBI observations between 2003 and 2005 by \citet{ulvestad09}, where he reports that he finds no evidence for significant variability at the $\sim$20\% level. However, the individual flux densities at the various epochs of observation are not listed in his paper, which makes a detailed comparison with our data problematical. In fact, over 18 months a recent RSN may have gone from very low to very high flux density values, and even to have faded away completely.  Nevertheless, from his non-detection of A15 at 2.3 GHz, and our clear detection with an increasing flux density over two epochs at 5.0 GHz, we suggest that A15 is a relatively recent, slowly-evolving RSN. This behaviour is very similar to that displayed by A0 (see below). Similarly, source A12 shows an increasing flux density at 5.0 GHz; and since it was also previously detected in 2003 and 2005, it too is likely to be a relatively recent and slowly-evolving RSN.  The nature of A6 is less clear.  \citet{ulvestad09} detected A6 only at 2.3 GHz. This detection, together with the sudden drop in its 5.0 GHz flux density between April and December 2008 may be interpreted as the result of a CCSN leaving its young radio supernova phase (when its radio emission is powered by interaction with the CSM) and now entering the supernova remnant phase (when it is powered by interaction with the ISM). We cannot exclude, however, the possibility that A6 is an X-ray binary or a microquasar.  In fact, A6 is quite close to two X-ray sources reported by \citet{zezas03} (sources 14 and 16 in their Table~\ref{tab,evn}).  The combined absorption corrected X-ray luminosity of those sources is $\sim 1.9 \EE{40}$\ergs.  From the 5.0 GHz radio luminosity of A6 in Table~\ref{tab,evn}, and assuming a spectral index of $\alpha$=-0.5 (typical of microquasars), the approximate total radio luminosity is $\sim4.8\EE{36}$\ergs, resulting in a ratio of radio-to-X ray luminosity of $\sim 2.5\EE{-4}$, which is a bit high, but still compatible with A6 being a microquasar.  Approved new EVN high-sensitivity observations will allow us to confirm the nature of A6 in the near future.

The maximum 5.0 GHz luminosities inferred by A12 and A15 are of $\sim$1.6$\EE{27}$\ergshz\/ and $\sim$7.6$\EE{26}$\ergshz, respectively. These are typical values of radio emitting Type IIP, or Type IIb SNe \citep{chevalier06}.  If confirmed, it would be the first time that such relatively faint radio SNe have been detected in the nuclear starburst of a local (U)LIRG. For comparison, Arp 220 appears to contain essentially very bright radio supernovae, which are identified with Type IIn SNe \citep{parra07}. We note here that this is not simply a sensitivity issue, since some of the VLBI observations of Arp 220 had r.m.s. as low as 9$\mu$Jy/b \citep{parra07}, yet most of the objects detected there were identified, based on their large radio luminosities, as Type IIn SNe.  We cannot exclude, however, that A12 and A15 have not yet reached their peak. In that case, their (peak) radio luminosity could be a few times higher, and given their slow evolution, these SNe could be Type IIL, or even Type IIn.

 In a previous paper, Neff et al. (2004) reported the detection of five VLBI sources (A0 to A4), within the central $\sim$80 pc of Arp 299-A. We detect all of these components at 5.0 GHz.  Components A1 to A4 do not show significant variability in the eight months covered by our 5.0 GHz eEVN observations, which is consistent with them being young supernova remants, as is also suggested by \citet{ulvestad09}.  

 A0 was first detected at 8.4 GHz, and identified with a radio supernova by \citet{neff04}.  We have now detected it at 5.0 GHz, more than five years after its discovery at 8.4 GHz. This implies that A0 is a long-lasting, slowly-evolving, non-standard radio supernova, since most other examples evolve more rapidly \citep{weiler02}.  Although rather uncommon, there are similar cases reported in the literature, both in normal galaxies, e.g., SN 1979C in M100 \citet{montes00}, and in (U)LIRGs, like some of the RSNe in Arp 220 \citep{parra07}.  In addition, its non-detection at 2.3 GHz up to 2005 \citep{ulvestad09} suggests the presence of a foreground absorber (e.g., a nearby H~II region), as in the case of SN 2000ft in the LIRG NGC~7469 (Alberdi et al. 2006; P\'erez-Torres et al.  2009).  Finally, we note that A5 may also show similar properties, although its variability significance is lower. In short, the behaviour displayed by A0 (and A5) whilst unusual, is not unknown for an RSN, and provides important information on how the interaction between the SN and CSM is proceeding, and thus probes the mass-loss history of the progenitor star. We therefore suggest that it is the local CSM conditions which are primarily responsible for determining the power of the radio supernovae exploding in the nuclear starburst of Arp 299-A.

e-EVN observations show that Arp 299-A hosts an extremely prolific supernova factory,  with radio luminosities typical of Type IIb, IIp, and IIL, and  yielding strong support to the scenario proposed by 
\citet{almudena00}, which suggests the existence of a recent (less than 10$-$15 Myr old) intense, instantaneous starburst.  We find evidence for at least three slowly-evolving, long-lasting, non-standarnd RSNe (A0, A12, and A15), which is very suggestive of the local CSM conditions playing a main role in determining the radio behaviour of the exploding SNe.
Our current monitoring of Arp 299-A with the eEVN at 5.0 GHz, which is scheduled to continue until the end of 2010- should allow us to detect any new radio supernova,  and therefore test whether the IMF in Arp 299-A is top-heavy, in contrast with normally assumed Salpeter \citep{salpeter55}, or Kroupa \citep{kroupa01} IMFs, where the production of massive stars (M $\gsim$8 M$_{\odot}$) that eventually result in CCSNe is low compared to the production of less massive stars. There seems to be evidence that this  might also be the case of M~82 \citep{doane93} and Arp~220 \citep{parra07}, and theoretically it is expected that in the warm dense ISM conditions within a (U)LIRG, the IMF should indeed be top-heavy due to a larger Jeans mass \citep{klessen07}.


\begin{table*}
\centering                          
\caption{Compact radio emitting sources in the central region of Arp 299-A}             
\label{tab,evn}      
\renewcommand{\footnoterule}{}
\begin{tabular}{llccllrrr}
\hline \hline
Source & Source & $\Delta\,\alpha^{\mathrm{a}}$ & $\Delta\,\delta$ & \multicolumn{2}{c}{S$_\nu$ ($\mu$Jy)$^{\mathrm{b}}$} 
& \multicolumn{2}{c}{L$_\nu$/10$^{26}$~\ergshz} & $V^{\mathrm{c}}$ \\
\cline{5-8}
Name$^{\mathrm{d}}$  & Type & (J2000.0)  & (J2000.0)  & 8 Apr 2008  & 5 Dec 2008 & 8 Apr 2008  & 5 Dec 2008 & \\
 \hline 
A0 & SN       & 33.6212 & 46.707 & 318$\pm$42 & 446$\pm$33 &  7.9$\pm$1.0 & 11.1$\pm$0.8 & 2.4\\
A1 & SNR    & 33.6199 & 46.699 & 855$\pm$58 & 901$\pm$51 & 21.3$\pm$1.4 & 22.4$\pm$1.3 & 0.6\\
A2 & SNR      & 33.6219 & 46.655 & 708$\pm$53 & 713$\pm$43 & 17.6$\pm$1.3 & 17.7$\pm$1.1 & 0.1\\
A3 & SNR      & 33.5942 & 46.560 & 398$\pm$44 & 353$\pm$30 &  9.9$\pm$1.1 &  8.8$\pm$0.7 & 0.8\\
A4 & SNR      & 33.6501 & 46.537 & 558$\pm$48 & 628$\pm$40 & 13.9$\pm$1.2 & 15.6$\pm$1.0 & 1.1\\
A5 & SN?      & 33.6176 & 46.693 & 278$\pm$41 & 143$\pm$25 &  6.9$\pm$1.0 &  3.6$\pm$0.6 & 2.8\\
A6 & SN       & 33.6206 & 46.597 & 208$\pm$40 & $\leq$72   &  5.2$\pm$1.0 & $\leq$1.8    & 3.4\\
A7 & SNR      & 33.6300 & 46.786 & 496$\pm$46 & 468$\pm$34 & 12.3$\pm$1.2 & 11.6$\pm$0.8 & 0.5\\
A8 & SNR      & 33.6306 & 46.401 & 226$\pm$41 & 264$\pm$27 &  5.6$\pm$1.0 &  6.6$\pm$0.7 & 0.8\\
A9 & SNR      & 33.6306 & 46.620 & 294$\pm$42 & 282$\pm$28 &  7.3$\pm$1.0 &  7.0$\pm$0.7 & 0.2\\
A10& SNR      & 33.6392 & 46.551 & 550$\pm$48 & 436$\pm$32 & 13.7$\pm$1.2 & 10.9$\pm$0.8 & 2.0\\
A11& SNR      & 33.6403 & 46.583 & 300$\pm$42 & 351$\pm$30 &  7.5$\pm$1.0 &  8.7$\pm$0.7 & 1.0\\
A12& SN       & 33.6495 & 46.590 & 449$\pm$45 & 639$\pm$40 & 11.2$\pm$1.1 & 15.9$\pm$1.0 & 3.2\\
A13& SN?      & 33.6531 & 46.733 & 251$\pm$41 & 118$\pm$25 &  6.2$\pm$1.0 &  2.9$\pm$0.6 & 2.8\\
A14& SNR      & 33.6825 & 46.571 & 292$\pm$42 & 260$\pm$27 &  7.3$\pm$1.0 &  6.5$\pm$0.7 & 0.6\\
A15& SN       & 33.5991 & 46.638 & 159$\pm$40 & 304$\pm$28 &  4.0$\pm$1.0 &  7.6$\pm$0.7 & 3.0\\
A16& uncl.        & 33.6149 & 46.516 & $\leq$117 & 147$\pm$25 &  $\leq$2.9  & 3.7$\pm$0.6    & 1.2\\
A17& uncl.        & 33.6179 & 46.561 & $\leq$117 & 179$\pm$26 &  $\leq$2.9  & 4.5$\pm$0.6    & 2.4\\
A18& uncl.        & 33.6192 & 46.464 & 151$\pm$40 & 129$\pm$25 &  3.8$\pm$1.0 & 3.2$\pm$0.6  & 0.5\\
A19& SN?      & 33.6196 & 46.659 & $\leq$117 & 191$\pm$26 &  $\leq$2.9 & 4.8$\pm$0.6     & 2.8\\
A20& uncl.        & 33.6278 & 46.789 & $\leq$117 & 146$\pm$25 &  $\leq$2.9 & 3.6$\pm$0.6     & 1.2\\
A21& uncl.        & 33.6291 & 46.836 & $\leq$117 & 133$\pm$25 &  $\leq$2.9 & 3.3$\pm$0.6     & 0.6\\
A22& uncl.       & 33.6354 & 46.678 & 173$\pm$40 & 217$\pm$26 &  4.3$\pm$1.0  & 5.4$\pm$0.7 & 0.9\\
A23& uncl.        & 33.6360 & 46.560 & $\leq$117 & 137$\pm$25 &  $\leq$2.9 & 3.4$\pm$0.6     & 0.8\\
A24& uncl.        & 33.6361 & 46.764 & $\leq$117 & 166$\pm$25 & $\leq$2.9 & 4.1$\pm$0.6      & 2.0\\
A25& SN?      & 33.6524 & 46.701 & 132$\pm$40 & 209$\pm$26 &  3.3$\pm$1.0  & 5.2$\pm$0.7 & 1.6\\
\hline
\end{tabular}   
\begin{list}{}{}
\item[$^{\mathrm{a}}$] Coordinates are given with respect
  to $\alpha$(J2000.0)= 11:28:00.0000 and $\delta$ (J2000.0) =
  58:33:00.000, and were obtained from the 5 December 2008 image. The
  positions for those sources also detected in 8 April 2008 coincide
  within the errors ($\lsim$0.5 mas) with all of them.
\item[$^{\mathrm{b}}$]  The uncertainty in the reported flux density for the detected compact components corresponds to 1$\sigma$, where $\sigma$
  was determined by adding in quadrature the off-source r.m.s in each image and a 5\% of the local maximum, to
  conservatively account for possible inaccuracies of the eEVN calibration.
\item[$^{\mathrm{c}}$] We define the significance of the flux density
  variability between the two consecutive epochs as $V = \mid S_{\rm
      Dec} - S_{\rm Apr}\mid / \sqrt{\sigma_{\rm Dec}^2+ \sigma_{\rm
      Apr}^2}$, where $S_{\rm Apr}$ and $\sigma_{\rm Apr}$ ($S_{\rm
    Dec}$ and $\sigma_{\rm Dec}$) are the values in columns 5 and 6 (7 and 8), respectively.
\item[$^{\mathrm{d}}$] Source names are given in right ascension
  order, except for the five components reported  previously (A0 through A4) by \citep{neff04}.
\end{list}
\end{table*}

%
%

\onltab{2}{
\begin{table*}
\centering                          
\caption{Brightness temperatures$^{\mathrm{a}}$  of the VLBI sources in Arp 299-A}             
\label{tab,tb}      
\begin{tabular}{lllccllrr}
\hline \hline
Source & \multicolumn{4}{c}{8 April 2008}                  & \multicolumn{4}{c}{5 December 2008}\\
Name   & S$_\nu$ ($\mu$Jy) & $a$(mas) & $b$(mas)& $T_B$(K) & S$_\nu$ ($\mu$Jy) & $a$(mas) & $b$(mas) & $T_B$(K) \\
\hline
A0 &  318$\pm$42 & $\leq  2.8$& $\cdots$    & $\cdots$ & 446$\pm$33 &5.3 & 3.0 & 1.8$\EE{6}$ \\
A1 &  855$\pm$58 & $\leq  3.6$& $\cdots$    & $\cdots$ & 901$\pm$51 &2.1 & 1.8 & 1.6$\EE{7}$ \\
A2 &  708$\pm$53 & $\leq  2.5$& $\leq 1.5$  & $\geq 1.3\EE{7}$ & 713$\pm$43 & 2.1 & 1.6 & 1.4$\EE{7}$ \\
A3 &  398$\pm$44 & $\leq  2.2$& $\cdots$    & $\cdots$ & 353$\pm$30 &$\leq6.5$ & $\leq 0.3$ & $\geq 1.2\EE{7}$ \\
A4 &  558$\pm$48 & $\leq  3.6$& $\leq 2.3$  & $\geq 1.0\EE{8}$ & 628$\pm$40 &$\leq2.8$ & $\leq 0.5$ & $\geq 3.0\EE{7}$ \\
A5 &  278$\pm$41 & $\leq 11.1$& $\leq 1.2$  & $\geq 1.3\EE{6}$ & 143$\pm$25 & $\cdots$ &  $\cdots$ &  $\cdots$ \\
A6 &  208$\pm$40 & $\leq  7.6$& $\leq 1.4$  & $\geq 1.3\EE{6}$& $\leq$72   & $\cdots$ &  $\cdots$ &  $\cdots$ \\
A7 &  496$\pm$46 & $\leq  2.5$& $\leq 3.3$  & $\geq 4.0\EE{6}$& 468$\pm$34  & 3.4    & 2.0 & $4.6\EE{6}$ \\
A8 &  226$\pm$41 & $\leq  4.8$& $\cdots$    & $\cdots$& 264$\pm$27 & $\leq 3.3$ &  $\leq 0.7$ & $\geq 7.6\EE{7}$ \\
A9 &  294$\pm$42 & $\leq  3.6$& $\leq 4.4$  & $\geq 1.2\EE{6}$ & 282$\pm$28 & $\cdots$ &  $\cdots$ &  $\cdots$ \\
A10&  550$\pm$48 & $\leq  3.2$& $\leq 3.9$  & $\geq 2.9\EE{6}$ & 436$\pm$32 & $\cdots$ &  $\cdots$ &  $\cdots$ \\
A11&  300$\pm$42 & $\leq  5.5$& $\cdots$    & $\cdots$ & 351$\pm$30 &$\leq 4.5$ & $\leq 1.8$ & $\geq 2.0\EE{6}$ \\
A12&  449$\pm$45 & $\leq  3.1$& $\leq 3.7$  & $\geq 2.6\EE{8}$ & 639$\pm$40 &2.3 & 1.5 & 1.2$\EE{7}$ \\
A13&  251$\pm$41 & $\leq  3.3$& $\leq 4.2$  & $\geq 1.2\EE{6}$ & 118$\pm$25 & $\leq 5.3$ & $\cdots$ & $\cdots$ \\ 
A14&  292$\pm$42 & $\leq  5.0$& $\leq 3.7$  & $\geq 1.1\EE{6}$ & 260$\pm$27 & $\leq 4.9$ &  $\cdots$ &  $\cdots$ \\
A15&  159$\pm$40 & $\cdots$ &  $\cdots$ &  $\cdots$ & 304$\pm$28 & $\leq2.8$ & $\leq 1.6$ & $\geq 4.5\EE{6}$ \\
A16&  $\leq$117  & $\cdots$ &  $\cdots$ &  $\cdots$ & 147$\pm$25 & 5.6 & 3.2 & $5.5\EE{5}$ \\
A17&  $\leq$117  & $\cdots$ &  $\cdots$ &  $\cdots$ & 179$\pm$26 & 7.3 & 3.9 & $4.2\EE{5}$ \\
A18&  151$\pm$40 & $\cdots$ &  $\cdots$ &  $\cdots$ & 129$\pm$25 & $\leq 5.1$& $\leq 5.5$ & $\geq 3.1\EE{5}$ \\
A19&  $\leq$117  & $\cdots$ &  $\cdots$ &  $\cdots$ & 191$\pm$26 & $\leq 4.0$& $\leq 2.0$ & $\geq 1.6\EE{6}$ \\
A20&  $\leq$117  & $\cdots$ &  $\cdots$ &  $\cdots$ & 146$\pm$25 & $\leq 7.9$& $\leq 3.5$ & $\geq 3.5\EE{5}$ \\ 
A21&  $\leq$117  & $\cdots$ &  $\cdots$ &  $\cdots$ & 133$\pm$25 & $\leq 7.4$& $\leq 2.9$ & $\geq 4.1\EE{5}$ \\
A22&  173$\pm$40 & $\cdots$ &  $\cdots$ &  $\cdots$ & 217$\pm$26 & $\leq 8.6$& $\leq 4.3$ & $\geq 3.9\EE{5}$ \\
A23&  $\leq$117  & $\cdots$ &  $\cdots$ &  $\cdots$ & 137$\pm$25 & $\leq 8.8$& $\leq 2.9$ & $\geq 3.6\EE{5}$ \\
A24&  $\leq$117  & $\cdots$ &  $\cdots$ &  $\cdots$ & 166$\pm$25 & $\leq 8.6$& $\leq 7.0$ & $\geq 1.8\EE{5}$ \\
A25&  132$\pm$40 & $\cdots$ &  $\cdots$ &  $\cdots$ & 209$\pm$26 & $\leq 8.6$& $\leq 4.4$ & $\geq 3.7\EE{5}$ \\
\hline
\end{tabular}   
\begin{list}{}{}
\item[$^{\mathrm{a}}$] The brightness temperatures shown for the 5.0 GHz VLBI
source components in Arp 299-A were calculated using the flux densities
in Table~ \ref{tab,evn} and the angular sizes quoted here.  We derived the
brightness temperatures from the general formula: $T_{\rm
  B} = (2\,c^2/k)\,B_\nu\,\nu^{-2}$, where $B_\nu$ is the intensity,
in erg~s$^{-1}$~Hz$^{-1}$~str$^{-1}$.  Since $B_\nu$ depends on the measured
flux density, $S_\nu$, and on the deconvolved angular
size of each VLBI component (obtained by fitting them to 
elliptical Gaussians, characterized by their major and minor semi-axis, 
$a$ and $b$). Therefore, the above formula can be
rewritten as $T_{\rm B} = (2\,c^2/k)\,B_\nu\,\nu^{-2} =
1.66\EE{9}\,\,S_\nu \, \nu^{-2} (a\,b)^{-1}$, with $S_\nu$ in mJy, $\nu$ in GHz,
and $a$ and $b$ in milliarcseconds, respectively.
\end{list}
\end{table*}   
}

\begin{acknowledgements}
We are very grateful to the anonymous referee for many suggestions and comments which have significantly improved the science and contents of our paper.
  We also thank L. Colina, A. Alonso-Herrero, J.M. Torrelles, E. Alfaro, and S. Mattila for many useful comments on the manuscript and insightful discussions. 
The continuing development of e-VLBI within the EVN is made possible via the EXPReS project funded by the EC FP6 IST Integrated infrastructure initiative contract \#026642 - with a goal to achieve 1 Gbit/s e-VLBI real time data transfer and correlation.
The EVN is a joint facility of European, Chinese, South African and other radio
  astronomy institutes funded by their national research councils.
  MAPT, CRC, and AA acknowledge support by the Spanish Ministry of
  Education and Science (MEC) through grant AYA 2006-14986-C02-01.
\end{acknowledgements}

\bibliographystyle{aa}
\bibliography{miguel-biblio}

\Online

\begin{appendix}
\section{eEVN observations of Arp 299-A}
\label{app,evn}

\subsection{Observing strategy}

We observed the central regions of Arp 299-A  at a frequency of 5~GHz in two epochs using the EVN.  Our first epoch on 8-9 April 2008 (2008.99; experiment code RP009) included the following six antennas (acronym, diameter, location): Cambridge (CM, 32~m, United Kingdom), Medicina (MC, 32~m, Italy), Jodrell Bank (JB, 76~m, United Kingdom), Onsala (ON, 25~m, Sweden), Torun (TO, 32~m, Poland), and Westerbork array (WB, 25~m, The Netherlands). Our second observing epoch on 5 December 2008 (2008.340; RP014A experiment) included, in addition the EVN antennas Effelsberg (EF, 100~m, Germany), Knockin (KN, 25~m, United Kingdom), and Shanghai (SH, 25~m, China).

Both observing epochs consisted of e-VLBI phase-referenced experiments, using a data recording rate of 512~Mbps with two-bit sampling, for a total bandwidth of 64~MHz.  The data were correlated at the EVN MkIV Data Processor at JIVE using an averaging time of 1~s. First epoch observations consisted of $\sim$8.0 hr on target. The telescope systems recorded both right-hand and left-hand circular polarization (LCP and RCP) which, after correlation, were combined to obtain the total intensity mages presented in this paper.  4.5 minute scans of our target source, Arp 299-A, were alternated with 2 minute scans of our phase reference source, J1128+5925. 3C345 and 4C39.25 were used as fringe finders and band-pass calibrators.  Our second epoch consisted of $\sim$4.5 hr on target. The telescope systems also recorded in dual polarization, and 4.5~minute on-source scans were alternated with 1 minute scans of J1128+5925. The bright sources 3C84, 3C138, 4C39.25 and 3C286 were used as fringe finders and band-pass calibrators.  (We note that the inclusion of EF in the second observing epoch allowed us to get a much better r.m.s., and this in spite of the significantly less amount of total observing time.)

\subsection{Data calibration and imaging}

We analyzed the correlated data for each epoch using the NRAO Astronomical Image Processing System ({\it AIPS}; http://www.aips.nrao.edu).  The visibility amplitudes were calibrated using the system temperature and gain information provided for each telescope. Standard inspection and editing of the data were done within AIPS.  The bandpasses were corrected using the bright calibrator 4C39.25.  We applied standard corrections to the phases of the sources in our experiment, including ionosphere corrections (using total electron content measurements publicly available).

Due to the limited bandwidth of CM and KN, the usable data of these antennas was found in a single sub-band with very noisy edges.  To improve the quality of the bandpass calibration, we removed the edges of CM for the first epoch, and the edges of CM and KN for the second epoch.  The instrumental phase and delay offsets among the 8-MHz baseband converters in each antenna were corrected using a phase calibration determined from observations of 4C\,39.25.  The data for the calibrator J1128+5925 were then fringe-fitted in a standard manner.  We then exported the J1128+5925 data into the Caltech imaging program DIFMAP \citep{shepherd95} for mapping purposes.  We thus determined gain correction factors for each antenna.  J1128+5925 showed a flux of $\sim 0.38$\,Jy at 5.0 GHz at both epochs, and displayed a point-like structure at mas scales.
After this procedure was completed within DIFMAP, the data were read back into AIPS, where the gain corrections determined by DIFMAP were applied to the data.

The final source model obtained for J1128+5925 was then included as an input model in a new fringe-fitting search for J1128+5925, thus removing the structural phase contribution to the the solutions of the delay and fringe rate for our target source, Arp 299-A, prior to obtaining the final eEVN images shown in Figure \ref{fig,arp299}.  The phases, delays, and delay-rates determined for J1128+5925 were then interpolated and applied to the source Arp 299-A. This procedure allowed us to obtain the maximum possible accuracy in the positions reported for the compact components in Figure \ref{fig,arp299}.  We note, though, that this has no impact in the final images, since J1128+5925 is essentially point-like at the angular resolution ($\lsim$9 milliarcseconds) provided by our 5.0 GHz e-EVN observations, and therefore its phase-contribution is negligible.

The imaging and deconvolution of Arp 299-A was done with the AIPS task IMAGR (see Figures \ref{fig,arp299} and \ref{fig,arp299-bphlame}), using a natural weighting scheme and a ROBUST parameter equal to zero. This scheme was a good compromise to maximize a very high angular resolution and a very low r.m.s. values in our images.  We point out that we kept the averaging integration time to 1\,s and used a maximum channel bandwidth in the imaging process of 8\,MHz, which results in a maximum degradation of the peak response for the component furthest away from the phase center of less than 5\% and prevents artificial smearing of the images (e.g., \citealt{bridle99}).

No self-calibration of the phases, nor of the amplitudes, was performed on the data, since the peaks of emission were very faint for such procedures.  We emphasize here two powerful aspects of the use of the phase-reference technique: (i) it effectively increases the integration time on a source from minutes to hours, thus increasing significantly the array sensitivity, and allowing the detection and imaging of very faint objects \citep{beasley95} and (ii) it retains the positional information of the compact components in Arp 299-A with respect to the phase-reference source J1128+5925, which lies 0.86$^\circ$ from the target source.

\subsection{Source detection, identification, and flux density extraction}
\label{app,detection}

Our 5.0 GHz eEVN observations on 8 April 2008 and 5 December 2008 resulted in off-source r.m.s noises of 39$\mu$Jy/b and 24$\mu$Jy/b, respectively. The difference in sensitivity was primarily due to the addition of the Effelsberg antenna in our second observing epoch, but also to the fact that there were more antennas than in our first epoch. We considered as real sources those individual peaks which had a signal-to-noise ratio equal to, or above five times the r.m.s. noise in either of our two observing epochs.  Since our second observing epoch was much more sensitive than the first one, this resulted in a larger number of detections of compact components (26) with respect to the first epoch (15).  We then put small boxes in the 8 April 2008 image, around the positions of those components detected on 5 December 2008, to extract the flux density of the counterparts initially detected only in the second epoch (components A15 through A25 in Figure \ref{fig,arp299}).  This procedure allowed us to recover four components above 3$\sigma$ (A15, A18, A22, and A25), based on a positional coincidence with the peak of emission in our second epoch to better than 0.5 milliarcsec; and for those components not detected on 8 April, to derive 3$\sigma$ upper limits to the emission.

The flux density extraction was carried out within AIPS, using task IMFIT, which fits Gaussian components to the image. Namely, we put small boxes (the few inner pixels of each source) around each local peak of brightness (above five times the r.m.s. noise) and fitted single Gaussian components to each of them.  Since the fitted integrated flux densities did not differ significantly from the peaks of emission (less than 1\%), it is clear that there is no evidence that any of the compact components are resolved even with the very high angular resolution of a few milliarcsec that the eEVN delivers. The total flux densities which are listed in Table~\ref{tab,evn} are thus the peak flux densities found in our images.

\begin{figure} \centering
  \includegraphics[width=90mm,angle=0]{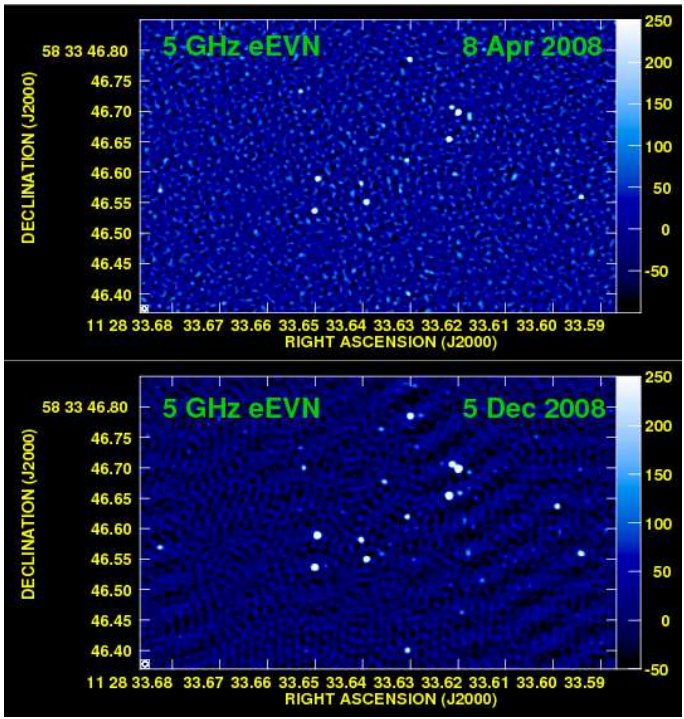} \\
  \caption{ {\it  5.0 GHz eEVN observations of the central $\sim$150 pc years of the Luminous Infrared galaxy Arp 299-A on 8 April 2008 (top) and 5 December 2008 (bottom), which reveals a large population of relatively bright, compact, non-thermal emitting sources. The colour scale goes from -50$\mu$Jy/b up to 250 $\mu$Jy/b. The size of
 the FWHM synthesized interferometric beam was of 
of 7.3 mas $\times$
6.3 mas and 8.6 mas $\times$ 8.4 mas, respectivetly. } }
 \label{fig,arp299-bphlame}
\end{figure}

\end{appendix}

\end{document}